\documentclass[12pt,preprint]{aastex}

\def\1h{{1H 0419-577}}

\def\xmm{{\it XMM-Newton}}

\def\et{{et al.\ }}

\def\rosat{{\it ROSAT}}

\def\euve{{\it EUVE}}

\def\asca{{\it ASCA}}
\def\sax{{\it BeppoSAX}}

\def\arcs{{\hbox{$^{\prime\prime}$}}}

\def\Msun{\hbox{$\rm ~M_{\odot}$}}

\def\H0{{\rm ~km~s^{-1}~Mpc^{-1}}}

\def\et{{et al.}}

\def\deg{^\circ}

\shorttitle{XMM-Newton observation of 1H0419-577}
\shortauthors{Pounds et al.}

\begin{document}


\title{An \xmm\ observation of the Seyfert 1 galaxy \1h\ in an extreme low state}


\author{K.A.Pounds\altaffilmark{1}, J.N.Reeves\altaffilmark{2,3}, K.L.Page\altaffilmark{1},  P.T.O'Brien\altaffilmark{1}}

\altaffiltext{1}{Department of Physics and Astronomy, University of Leicester, Leicester LE1 7RH, UK}
\altaffiltext{2}{Laboratory for High Energy Astrophysics, Code 662, NASA Goddard Space Flight Center,
Greenbelt, MD 20771, USA}
\altaffiltext{3}{Universities Space Research Association}
\email{kap@star.le.ac.uk}


\begin{abstract}

Previous observations of the luminous Seyfert 1 galaxy \1h\ have found its X-ray spectrum to range from that of a typical Seyfert 1
with 2--10 keV power law index $\Gamma$$\sim$1.9 to a much flatter power law of $\Gamma$$\sim$1.5 or less. We report here a new
\xmm\ observation which allows the low state spectrum to be studied in much greater detail than hitherto. We find a very hard
spectrum ($\Gamma$$\sim$1.0), which exhibits broad features that can be modelled with the addition of an extreme relativistic Fe K
emission line or with partial covering of the underlying continuum by a substantial column density of near-neutral gas. Both the
EPIC and RGS data show evidence for strong line emission of OVII and OVIII requiring an extended region of low density
photoionised gas in \1h. Comparison with an earlier \xmm\ observation when \1h\ was `X-ray bright' indicates the dominant spectral
variability occurs via a steep power law component.

\end{abstract}


\section{Introduction}

\1h (also known as LB 1727) is a radio-quiet (8.4~GHz flux $\sim$3~mJy; Brissenden \et 1987) Seyfert galaxy at a redshift
z~=~0.104 and one of the brightest AGN in the extreme ultra-violet, being detected by both the \rosat\ Wide Field Camera (Pye
\et\ 1995) and \euve\ (Marshall \et\  1995). 
Optical spectra from the AAT (Turner \et\ 1999) and ESO (Guainazzi \et\ 1998) showed \1h\ to be
a typical broad line Seyfert 1 with a strong `Big Blue Bump'. 

\1h\ was observed by \xmm\ in December 2000, for $\sim$8 ksec, although a command error meant X-ray data were restricted to the
EPIC pn-camera. Analysis of the X-ray spectrum found a high 2--10 keV luminosity of $\sim$$10^{45}$ erg
s$^{-1}$, with a `canonical' Seyfert 1 power law continuum of $\Gamma$ $\sim 1.9$, together with a strong soft X-ray
excess (Page \et\ 2002). However, simultaneous \rosat\ HRI and \asca\ (Turner \et\ 1999) and \sax\ (Guainazzi \et\ 1998)
observations of \1h\ four years earlier had found an unusually hard spectrum for a Seyfert 1 galaxy, with $\Gamma$$\sim$1.4,
and only a very weak `soft excess'. A still earlier \rosat\ observation of \1h\ , in 1992, showed the source had once more been in a
high/soft state (Guainazzi \et\ 1998). \1h\ therefore appears to exhibit unusually large spectral variability, with a
factor-of-10 change in soft X-ray flux and variations in power law index by $\Gamma$$\ga0.5$, over several years or less. In
this respect, \1h\ is similar to Galactic black hole sources, which frequently vary between low/hard and high/soft states (see
Zycki \et\ 2001 for a review). Although the spectral changes in \1h\ are not quite so extreme, the unusually large variability
makes it a particularly interesting probe of the accretion-driven processes that make AGN characteristically powerful X-ray
emitters.

The impression from the published observations of \1h\ is that the soft X-ray flux shows the most dramatic changes, much
greater than the integrated hard X-ray flux, suggesting the spectral change is driven by the thermal disc emission. In the
framework of the widely accepted disc/corona model (Haardt and Maraschi 1991), where thermal photons from the accretion disc
are Compton scattered by high energy electrons to form the observed X-ray power law, an enhanced disc emission should cool the
electrons and lead to a steeper power law.  Structural changes in the scattering corona may also be required by the
observed power law slope change of $\Gamma$$\ga0.5$, with the additional constraint that the total 2-10 keV luminosity
increased by only 20 percent over the same 1996-2000 period (Page \et\ 2002).  

In considering changes in the structure of the disc, one possibility is that the low/hard state corresponds to a truncated
inner disc, or ADAF flow (which would lead to reduced reflection, no broad Fe~K$\alpha$ line, and a weaker coupling between
the disc and the corona). In that simple picture a hot, `photon-starved' corona could evolve to a high/soft state when a
higher accretion rate extended the disc boundary inward, with enhanced thermal emission, a steeper power law and broad
Fe~K$\alpha$ line. That might describe the change occurring in \1h\ between 1996 and 2000 (though the sensitivity of none of
the contemporary observations was sufficient to usefully constrain the Fe-K emission line; Guainazzi \et\ 1998,Turner \et\ 1999,  
Page \et 2002).

To improve the X-ray data on \1h\ a new series of 6 \xmm\ observations, at approximately 3-monthly intervals over the period
September 2002 to November 2003, was recently completed. The first of those new observations, when \1h\ was found to be extremely
faint, is reported in the present paper.

\section{Observations}

The first \xmm\ observation of \1h\ in the new series took place on 5 September 2002 (orbit 512) yielding a useful exposure of
$\sim$14.9 ksec. Five further observations of similar length were successfully carried out over the following 15 months and
will be reported later. The present paper describes the results from the first new observation, which are of particular
interest given the extremely low flux-state in which \1h\ was found.  X-ray data were available throughout the observation from
the EPIC pn (Str\"{u}der \et 2001) and MOS (Turner \et\ 2001) cameras, and the Reflection Grating Spectrometer/RGS (den Herder
\et\ 2001). In addition the Optical Monitor (Mason \et\ 2001) obtained simultaneous flux measurements in V,B,U and 2
ultraviolet wavebands. 

The X-ray data were first screened with the XMM SAS v5.4 software and events corresponding to patterns 0-4 (single and double
pixel events) were selected for the pn data and patterns 0-12 for MOS1 and MOS2, the latter then being combined. A low energy
cut of 300 eV was applied to all X-ray data and known hot or bad pixels were removed. Source counts were obtained from a
circular region of 45\arcs\ radius centred on \1h\ , with the background being taken from a similar region offset from, but
close to, the source. The X-ray light curve of \1h\ was essentially flat throughout the observation and the background rate was
low. We therefore integrated the total data set for spectral analysis. Individual EPIC spectra were binned to a minimum of 20
counts per bin to facilitate use of the $\chi^2$ minimalisation technique in spectral fitting.  RGS data were initially
grouped in sets of 10 spectral bins to aid the detection of weak features. Spectral fitting was based on the Xspec package
(Arnaud 1996) and all fits included absorption due to the line-of-sight Galactic column $N_{H}=2\times10^{20}\rm{cm}^{-2}$.
Errors are quoted at the 90\% confidence level ($\Delta \chi^{2}=2.7$ for one interesting parameter).

\section{EPIC spectrum} 

Figure 1 shows the overall shape of the EPIC 0.3--10 keV spectrum, compared with a simple power
law fit, with photon index $\Gamma$$\sim$1.49. In addition to the usual (but more extreme) 
curvature found in such a fit to
Seyfert 1 spectra, sharp features are seen in both pn and MOS data at $\sim$0.5 keV and $\sim$6 keV. The former feature
lies close to the neutral oxygen edge in the instrument response, but also coincides with potentially strong OVII and OVIII
emission lines (a possibility we check in Section 4 with the higher resolution RGS data). The high energy spectral feature is
suggestive of a strong, relativistically broadened Fe K emission line or a deep Fe K absorption edge. We test those alternatives
below.

\subsection{2--10 keV spectral fit with a Laor line and continuum reflection}

We began our spectral analysis of the EPIC data by fitting a power law over the hard X-ray (2--10 keV) band, hoping thereby to
exclude the effects of soft X-ray emission and/or low energy absorption. This fit yielded an extremely flat power law, with
photon index of $\Gamma$$\sim$1.06 (pn) and $\Gamma$$\sim$1.02 (MOS). Statistically the simple power law fit over the 2--10 keV
band was quite good, with $\chi^{2}$ of 692 for 626 degrees of  freedom (dof). However, the spectral feature below $\sim$6 keV
is clearly seen in both pn and MOS data (figure 2). The addition of an absorption edge improved the fit ($\chi^{2}$ of 617/622
dof), with an edge energy (in the AGN rest-frame) of 7.10$\pm$0.06 keV (pn) and 7.07$\pm$0.07 keV (MOS). The coincidence of
this value with the K-edge of neutral Fe is a clear indicator of strong reflection or line-of-sight absorption in `cold'
matter. However, the power law plus absorption edge fit still left substantial curvature in the data:model residuals. 

Since the spectral curvature in the 2--6 keV band is reminiscent of an extreme relativistic Fe K emission line, a Laor
emission line (Laor 1991) was then added to the power law continuum (without the added absorption edge). An excellent fit was then
obtained ($\chi^{2}$ of 589 for 620 dof), for a (rest frame) line energy of 6.4$\pm$0.3 keV, disc emissivity index
$\beta$$\sim$4.5 and inner radius r$_{in}$$\sim$1.5 r$_{g}$, where r$_{g}$ is the gravitational radius. The disc inclination
was 44$\pm$4$\deg$ in this fit. In addition to being extremely broad the required line equivalent width (EW) was also large,
at $\sim$0.9 keV in both pn and MOS spectral fits. The underlying power law slope was essentially unchanged by the addition of the
Laor emission line.

Since the Laor line arises by reflection (implicitly from the inner accretion disc) it must be accompanied by strong continuum
reflection in a physically realistic fit. Given the above absorption edge fit lies close to 7.1 keV (the K-edge energy of neutral Fe)
we chose to model the continuum reflection with PEXRAV in Xspec (Magdziarz and Zdziarski 1995), setting the high energy cut-off at 150 keV, with solar abundances and
the reflection factor R initially at 4 (to match the high Laor line EW). The subsequent best-fit, with R set free, was again very good
($\chi^{2}$ of 584 for 619 dof), with the power law slope increased by $\Delta$$\Gamma$$\sim$0.34 in both pn and MOS (normalisation
$\sim$$5.3\times10^{-4}$ ph cm$^{-2}$ s$^{-1}$),
and a reflection factor R=3.5$\pm$1.5. The Laor line parameters were little changed by the addition of the continuum reflection,
the line being still broad and strong.

\subsection{Extending the spectral fit to 0.3 keV}

Fixing the above 2--10 keV spectral parameters and extrapolating the fit to 0.3 keV showed a shallow deficit of flux between
$\sim$1--2 keV and a strong excess below $\sim$1 keV. The addition of a blackbody component of kT $\sim$90 eV modelled the
soft excess quite well, but it was necessary to free the power law and Laor line parameters to fit the data at 1--2 keV. A flattening
of the power law indices by $\sim$0.14 and an increase in the Laor line EW by $\sim$200 eV yielded an overall (0.3-10 keV) $\chi^{2}$ of
1135 for 1045 dof. The narrow peak in observed flux at $\sim$0.5 keV (visible in figures 1 and 2) was then the main contributor to the
data:model residuals. The addition of a gaussian emission line to model this feature gave a further significant improvement to the
broad-band fit, which was then formally acceptable ($\chi^{2}$ of 991 for 1039 dof), for a narrow line ($\sigma$=40$\pm$20 eV)
at 0.61$\pm$0.01 keV (pn) and 0.57$\pm$0.01 keV (MOS), with an EW$\sim$50 eV. 

We illustrate the overall `Laor line/PEXRAV' model spectrum in figure 3 and list the parameters of the model in Table 1. 
While complex, this model does provide an
excellent fit to the data. We briefly discuss the physical implications in section 6.1.  

The good match of model and data allows the mean X-ray fluxes and luminosity of \1h\ during the September 2002 \xmm\ observation to be 
derived. These were $1.9\times10^{-12}$~erg s$^{-1}$ cm$^{-2}$ (0.3--1 keV), with $\sim$67 percent in the blackbody component,
$0.9\times10^{-12}$~erg s$^{-1}$ cm$^{-2}$  (1--2 keV), and $8\times10^{-12}$~erg s$^{-1}$ cm$^{-2}$ (2--10 keV). Combining
these fluxes yields a 0.3--10 keV luminosity for \1h\ in the `low state' of $2.3\times 10^{44}$~erg s$^{-1}$ ($ H_0 = 75
$~km\,s$^{-1}$\,Mpc$^{-1}$).

\subsection{Partial covering fit}

Several recent studies (eg. Inoue and Matsumoto 2003, MCG-6-30-15; Pounds \et\ 2003a, PG1211+143) have pointed out that partial covering of the power law continuum
by absorbing matter can impose spectral curvature over the $\sim$3--7 keV band very similar in appearance to a relativistic Fe K
emission line. To test this alternative fit for the low state EPIC data of \1h\ we next considered a model in which a fraction of the
power law continuum is obscured  by an ionised absorber, using the ABSORI model in Xspec. The outcome was that both the 3--6 keV spectral
curvature and the absorption edge (observed at $\sim$6.4 keV, but at $\sim$7.1 keV in the AGN rest frame) were well fitted with 
60$\pm$15 percent of the power law being covered by weakly ionised matter of ionisation parameter $\xi$(= $L/nr^2$)$\leq$0.3
erg cm s$^{-1}$ and column density $N_{H}$= 4.3$\pm$0.4$\times10^{22}$~cm$^{-2}$. The power law index in this fit was less
extreme, with $\Gamma$$\sim$1.48 (pn) and $\Gamma$$\sim$1.36 (MOS). While these partial covering parameters are not unique, non-solar
abundances and a range of ionisation parameter allowing alternative fits to the spectral curvature, the observed absorption edge near 7 keV
requires the mean ionisation state to be low. Higher resolution spectra will be required to further constrain such absorbing matter.

Unlike for the Laor line/PEXRAV model, extrapolation of the 2--10 keV partial covering model remained a good fit to $\sim$1 keV, below which a blackbody
component was required to model the strong soft excess. Once again, this broad-band continuum fit left a significant excess of flux 
near $\sim$0.5 keV. Modelling this feature with a gaussian emission line produced similar parameters to those found in section 3.2,
giving a 0.3--10 keV broad-band fit of very similar overall quality ($\chi^{2}$ of 997 for 1038 dof). In this partial covering fit
it is interesting to note the data:model residuals now show an excess just below 6 keV. Adding, finally, a gaussian emission
line to the partial covering model gave a further small,
but statistically significant, improvement of $\Delta$$\chi^{2}$$\sim$16 for 4 additional parameters. The full details of the partial
covering fit are listed in Table 2, and the model is illustrated in figure 4. We discuss the physical implications of this model
and compare it with the Laor line/PEXRAV model in section 6.1.

To clarify the $\sim$0.5 keV emission feature seen in both the above fits to the EPIC spectra, and search for other structure in 
the soft X-ray spectrum of \1h\ , we then
examined the simultaneous RGS data.

\section{Spectral lines in the RGS data}

Both Laor and PC fits to the EPIC data indicate a strong soft excess above an extension
of the hard power law, and a narrow emission feature at $\sim$0.55-0.65 keV.

To gain further insight on the soft X-ray spectrum we examined the simultaneous \xmm\ grating data of \1h. We began by
jointly  fitting the RGS-1 and RGS-2 data with a power law and black body continuum (from the corresponding EPIC 0.3--10 keV
fits) and examining the data:model residuals by eye. The only obvious features (figure 5) were emission lines observed at
$\sim$21 $\AA$ and $\sim$24 $\AA$.  To quantify these features we then added gaussian emission lines to the power law plus blackbody
continuum fit in Xspec, with wavelength, line width and flux as free parameters. In each case the line width is
unresolved, indicating a FWHM$\leq$2000 km s$^{-1}$, though this value is not well constrained given the low RGS
count rates. The fitted spectrum is shown in figure 6.

In the rest frame of \1h\ the line wavelengths are 22.02$\pm$0.05 $\AA$ and 18.99$\pm$0.05 $\AA$, corresponding  closely
to the laboratory wavelengths of the forbidden OVII line (22.095 $\AA$) and the resonance OVIII Ly$\alpha$ line (18.969 $\AA$). The fitted line 
fluxes
were $6.1\times10^{-5}$~ph s$^{-1}$ cm$^{-2}$ and $7.7\times10^{-5}$~ph s$^{-1}$ cm$^{-2}$, respectively, corresponding to
equivalent widths of 18 and 28 eV. When blended in the lower resolution EPIC data these 2 strong emission lines match well
with the gaussian emission feature required in both pn and MOS spectral fits. The statistical quality of the joint RGS fit was
significantly improved by the addition of the two lines, with a reduction in $\chi^{2}$ of 36 for 4 fewer dof. 

Although the constraints on other emission (or absorption) lines in the RGS data are weak, due to the low flux state of
\1h, the dominance of the forbidden line of the OVII triplet is a clear signature of a photoionised plasma, with an electron
density $\leq$$10^{10}$cm$^{-3}$ (Porquet and Dubau 2000). The equally strong OVIII Ly$\alpha$ emission indicates an 
ionisation
parameter $\xi$ (= $L/nr^2$, where n is the gas density at a distance r from the ionising source of luminosity L) of order
45$\pm$10 (Kallman and McCray 1982).  Assuming a solar abundance of oxygen, with 40 percent in OVII, 50 percent of
recombinations from OVIII direct to the ground state, and a low temperature recombination rate of $10^{-11}$~cm$^{3}$~s$^{-1}$
(Verner and Ferland 1996), we deduce an emission measure for the forbidden line flux of order $2\times$$10^{66}$ cm$^{-3}$.

A further constraint on this ionised emission region can be obtained from the relevant ionisation parameter. Assuming a
`typical' (mean \1h\ flux) ionising luminosity for K-shell oxygen of L$_{OX}$$\sim$$2\times10^{44}$ erg s$^{-1}$, with
$\xi$$\sim$45, we find nr$^{2}$$\sim$$4\times10^{42}$ cm$^{-1}$. Combining this value with the above emission measure then
gives r $\sim$$3\times10^{19}$ cm for a uniform spherical distribution of photoionised gas, and a particle density n
$\sim$$8\times10^{3}$ cm$^{-3}$.  Such a low density plasma would have an equilibration time of $\sim$1 year, supporting the
above assumption of a flux-averaged ionising luminosity. In our analysis of the later \xmm\ observations of \1h\ it will be
instructive to see how the strong line emission from OVII and OVIII varies, if at all, noting also the light travel time
across a region as large as we estimate will be $\sim$30 years. 

\section{Comparison with the earlier high flux \xmm\ observation}  

A detailed analysis of the large-scale spectral variability in \1h\ should be possible by combining the data from all 6 \xmm\
observations made over the 15-month period from September 2002. The scale of the broad-band variability may already be indicated,
however, by comparing the present low flux state spectrum with the previously published pn observation from December 2000
(Page \et\ 2002). Figure 7 illustrates the degree of variability by comparing the ratio of the December 2000 data
to the best-fit 2--10 keV power law ($\Gamma$=1.9), with the ratio of the September 2002 data to the same power law.
The two data sets are seen to be quite similar at the highest energies, while diverging strongly below $\sim$5 keV.

To assess the gross features of the spectral variability we then obtained the difference spectrum of the two background-subtracted
pn data sets (adjusted for the different exposures) and compared the resulting data with a simple power law. The
fit was surprisingly good, with a power law of $\Gamma$=2.45$\pm$0.1 modelling the difference spectrum closely from 0.3--2 keV (figure
8). To check the apparent steepening of the power law at higher energies we re-grouped the data to a minimum of 500 bins, to
ensure adequate statistics in the highest energy data points. Fitting the difference spectrum above 2 keV then confirmed a
significant  steepening in the spectral slope, to $\Gamma$=2.7$\pm$0.1 (2--10 keV).

That comparison of the new `low state' EPIC spectrum with the earlier `high state' spectrum shows rather unambiguously that the
large scale spectral variability in \1h\ is dominated by a steep power law component. Furthermore, the steep power law fit to the
difference spectrum, $\Gamma$$\sim$2.45, implies that no `separate' variable soft emission (blackbody) component is required. The
gradual steepening of the difference spectrum at higher energies appears to be real and may offer an important clue to the
physical origin of the variable component. One obvious candidate is Comptonisation and a trial fit with compTT in Xspec (Titarchuk
1994) gave an excellent fit ($\chi^{2}$ of 101 for 105 dof), with input photons of kT$\sim$70 eV and an optically thick
Comptonising plasma ($\tau$=4.5$\pm$1) at kT=2.6$\pm$0.9 keV reproducing the mean slope and high energy downturn in the difference
spectrum.  

Future analysis of the full \xmm\ data set should shed further light on this interesting outcome.

\section{Discussion}
There are three points of particular note resulting from the \xmm\ observation of \1h\ reported here. 

Particularly remarkable is the extremely hard (flat) power law spectrum that approximates to the EPIC data over the 2--10 keV band. A
spectral index $\Gamma$$\sim$1.0 is flatter than any reported previously for this highly variable source, lying below all previous X-ray
spectra  (see fig.3 in Page \et\ 2002). In comparison we recall the $\sim$2--10 keV continua of radio quiet AGN have a photon index
usually in the range $\Gamma$=1.7--2.0 (Nandra and Pounds 1994, Reeves and Turner 2000).
Second, although the RGS features are relatively faint, the unambiguous detection of emission lines of OVII and OVIII provides clear
evidence for an extended region of photoionised gas in the nucleus of \1h.
Finally, a comparison of the raw EPIC data with data obtained in December 2000, when \1h\ was considerably brighter, gives a
model-independent indication that the large-scale spectral variability in \1h\ is primarily due to a variable, steep power law component.

\subsection{Relativistic Fe K line or partial covering?} 

However modelled, the 2-10 keV spectrum of \1h\ observed in September 2002 was very unusual. Although the addition of strong reflection,
or of partial covering, allowed for a steeper underlying power law continuum, it remained sufficiently flat to require rather extreme
conditions for Comptonisation models ( Svensson 1994, Haardt \et\ 1997), suggesting in particular a `photon starved' scenario'. Although
still to be proven by the analysis of further observations, the simple form of the difference spectrum ( obtained by subtracting the
present EPIC spectrum from that of December 2000) suggests the hard `low state' spectrum remained `constant' over that 21 month
interval.  The implication might be for a `core' accretion disc component that is located in a region of high gravity (the
reflection-dominated Laor line fit), or is overlain by substantial cool absorbing matter. 

Our finding that extreme relativistic Fe K emission line and partial covering models fit the spectral curvature in the $\sim$3--7 keV
band equally well adds to a growing number of such cases, most recently the low luminosity Seyfert NGC 4051 (Pounds \et\ 2003c). Both
models raise obvious questions in applying to a low flux state as seen here for \1h. In the former case there is the concern that strong
illumination of the innermost accretion disc, required to explain the extreme relativistic broadening of the Fe K emission line, is
counter-intuitive when the X-ray luminosity is apparently so low ( but see Miniutti and Fabian 2003), while in the latter the partial
covering requires a structured absorber covering the hard X-ray source, but not the region of soft X-ray emission. Statistically, our
fits to the EPIC data are equally good and - once again - higher energy data is needed to discriminate between the two models. Although
not conclusive, it is interesting to note that extending the EPIC fits to 12 keV offers some support for the PC model, where the
data:models residuals remain small (figure 4), whereas for the Laor line/PEXRAV model the flatter model continuum significantly
over-predicts the data (figure 3).

Further circumstantial evidence in favour of the PC model fit may be taken from the detection, in that case, of a non-relativistic Fe K
emission line. The line is weak, having an EW (against the total power law continuum) of only $\sim$85 eV, and hence poorly constrained.
Although the line energy is only barely compatible with fluorescence from neutral iron, the partial covering model does offer a natural
origin for an Fe K emission line of that order, by continuum absorption and fluorescent re-emission from the substantial column of
overlying gas (Makishima 1986). Incidentally, that explanation would leave little room for the narrow 6.4 keV line found to be a
prominent feature in many lower luminosity Seyfert galaxies ( and believed to arise from matter distant from the black hole, eg a
molecular torus), consistent with an emerging view that the EW of the narrow Fe K line is anti-correlated with luminosity (Page \et\
2003).

\subsection{Extended photoionised gas in \1h} 

Although only 2 emission lines were clearly detected in the RGS spectrum of \1h, taken together they provide a surprisingly  powerful
diagnostic of the emitting gas. Most useful is the detection of the forbidden line of OVII, since its relative strength (in the OVII
triplet) shows photoionisation to be the dominant process and sets a limit on the plasma density, while the measured line flux provides a
straightforward estimate of the emission measure. The detection of a similarly strong OVIII Ly$\alpha$ resonance line (usually seen in
absorption in Seyfert 1 galaxies, and perhaps enhanced here due to the low continuum flux of \1h\ ), allows the ionisation parameter of
the photoionised gas to be calculated, assuming a common location of the OVII and OVIII gas. The emission measure and ionisation
parameter then  provide estimates of the scale ($3\times$$10^{19}$ cm) of a spherical emission region of particle density
$\sim$$8\times10^{3}$ cm$^{-3}$. A corresponding mass for this extended gas envelope is then $8\times10^{5}$\Msun. Although evidence that
this gas is associated with an outflow must await a search for the corresponding absorption lines in the later \xmm\ observations of \1h,
a typical (Seyfert 1) outflow velocity of 300 km s$^{-1}$ (eg Kaspi \et\ 2002) could replenish this region in $\sim$$3\times10^{4}$
years.

\subsection{Spectral variability}

Comparison of the EPIC pn data from the observation reported here, with that from a much brighter state of \1h\ in December 2000,
has yielded some fascinating indicators. The large-scale change in the X-ray spectrum can be described by a variable intensity,
fixed slope, power law, which is sufficiently steep as not to require a separate variable soft (blackbody) component. This is
similar to the conclusion of Fabian \et\ (2003) from the long \xmm\ observation of MCG-6-30-15, though in that case the index of
the variable power law component was lower. We also note, if the hard power law component does remain unchanged (as implied
by the simple form of the difference spectrum), then it seems reasonable to conclude that the corresponding emission mechanism and
location are physically distinct from the variable power law component. 

If the variable X-ray component arises by Comptonisation of accretion disc photons, the scale of spectral change between the two
\xmm\ observations may be sufficiently great to require a structural change in the inner accretion disc and/or corona. We have some
evidence that the thermal disc emission was significantly stronger during the December 2000 observation, since the the OM UVW1
channel (the only one live) was 0.9 mag brighter than on the second occasion. Again, it is interesting to note that the variable
power law component has a spectral index ($\Gamma$$\sim$2.5) similar to that predicted for scattering in an optically thick
($\tau$$\sim$1) corona (Haardt \et\ 1997), suggesting the change is mainly in the scattering medium. Our simple compTT fit to the
difference spectrum (section 5) is consistent with that interpretation. 

In addition to direct or up-scattered disc radiation, a significant emission component might be associated with an energetic outflow, as
recently found in several luminous AGN (Chartas \et\ 2002, Pounds \et\ 2003a,b, Reeves \et\ 2003). While in those cases where high
velocity outflows have been confirmed they appear to be linked to a high (Eddington or super-Eddington) accretion rate (King and Pounds
2003), the evidence for column densities of highly ionised gas in excess of $N_H$$\sim$$10^{23}$ cm$^{-2}$ is becoming more common for
Seyfert 1 galaxies (eg Bianchi \et\ 2003).  The important point is that the kinetic energy in a high velocity flow can be comparable to
the accretion energy, offering the possibility - perhaps via shocks in the outflow - of an additional X-ray emission component. It is
interesting to speculate that the steep power law characterising the spectral variability in \1h\ might arise in this way, by providing
the additional Comptonising electrons or even by thermal emission. 

The extensive new \xmm\ observations of the large-scale spectral variability of \1h\ should shed new light on the
emission mechanism(s) which make AGN characteristically powerful X-ray sources. 

\section{Summary} 

A new \xmm\ observation of the luminous Seyfert 1 galaxy \1h\ has found the source to be in an extreme low flux state. The 2--10
keV spectrum is unusually hard, being approximated by a power law of $\Gamma$$\sim$1.0. However, significant residuals are
seen in the power law fit, which can be modelled by either a relativistic Fe K emission line or by partial covering by a
substantial column of near-neutral gas. Detection of emission lines of OVII and OVIII indicate the presence in \1h\ of an
extended region of highly ionised gas. A comparison of the present X-ray spectrum of \1h\ with that obtained in a short
observation by \xmm\ 2 years earlier, when the source was much brighter, shows that the spectral variability is dominated by a
previously unrecognised steep power law component.

\section{Acknowledgments}

The results reported here are based on observations obtained with \xmm, an ESA science
mission with instruments and contributions directly funded by ESA Member States and the USA (NASA). The authors wish to
thank the SOC and SSC teams for organising the \xmm\ observations 
and initial data reduction. KAP is pleased to acknowledge a Leverhulme Trust Emeritus Fellowship.

\clearpage

\begin{figure}
\rotatebox{-90}{
\epsscale{0.7}
\plotone{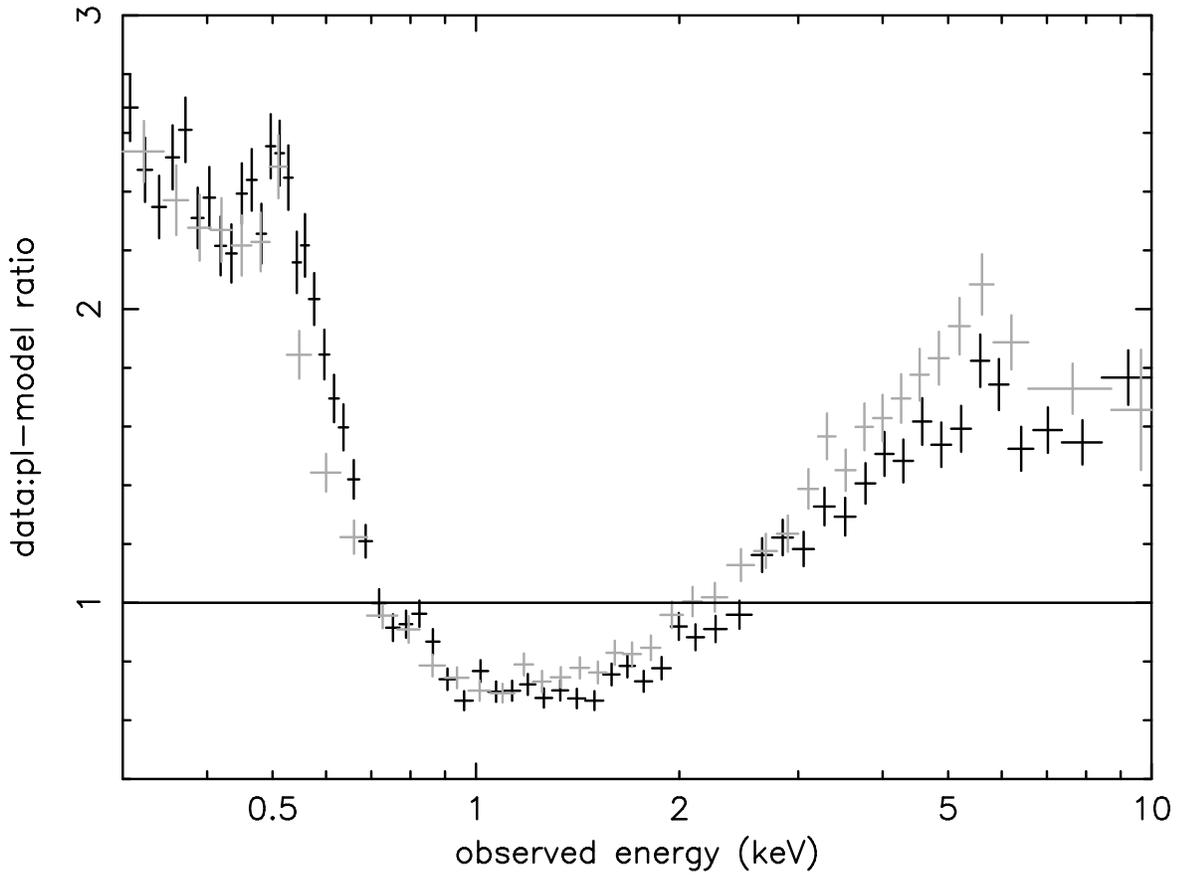}}
\caption{Ratio of data to a simple power law fit to the pn (black) and MOS (grey) EPIC spectra from the September 2002 
observation of \1h. 
\label{fig1}}
\end{figure}

\clearpage

\begin{figure}
\rotatebox{-90}{
\epsscale{0.7}
\plotone{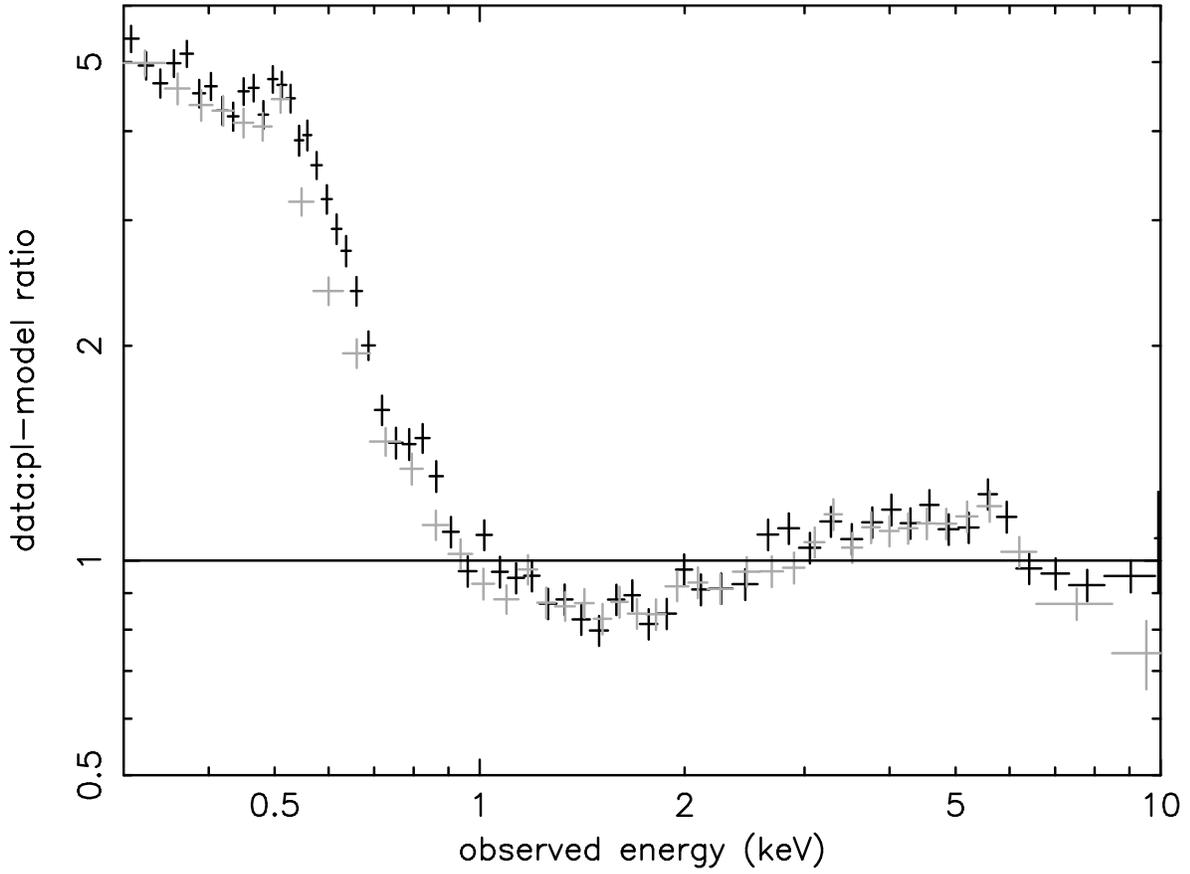}}
\caption{As figure 1 but with the power law fit restricted to the 2--10 keV band, showing the broad hump below $\sim$6 keV and strong
soft excess.\label{fig2}}
\end{figure}

\clearpage

\begin{figure}
\rotatebox{-90}{
\epsscale{0.7}
\plotone{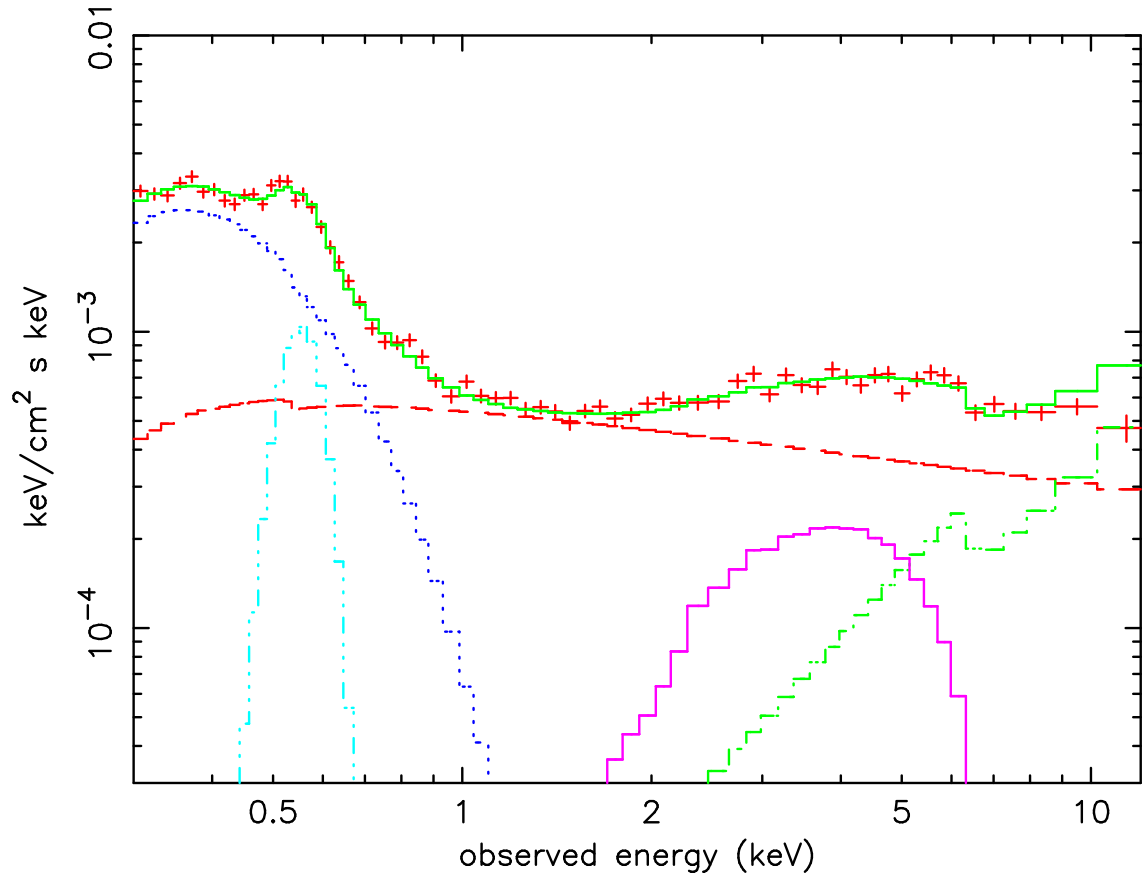}}
\caption{Unfolded model spectrum described in Section 3.2. Components are: power law (red) plus reflection (green), Laor line (pink),
blackbody (dark blue) and narrow gaussian emission line (pale blue).\label{fig3}}
\end{figure}

\clearpage

\begin{figure}
\rotatebox{-90}{
\epsscale{0.7}
\plotone{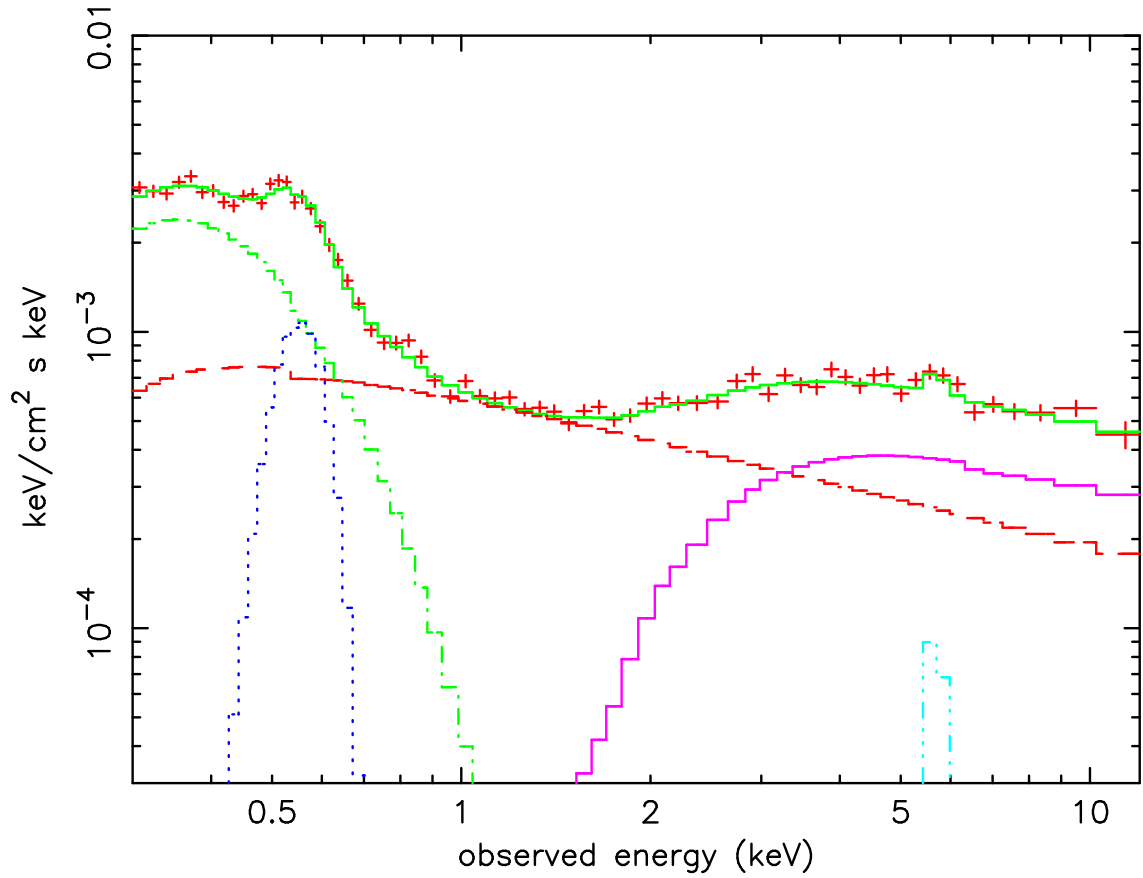}}
\caption{Unfolded model spectrum described in Section 3.3. Components are: partially-covered power law (red and pink), 
blackbody (green) and gaussian emission lines (blue) at $\sim$0.6 keV and $\sim$6.3 keV.\label{fig4}}
\end{figure}

\clearpage

\begin{figure}
\rotatebox{-90}{
\epsscale{0.7}
\plotone{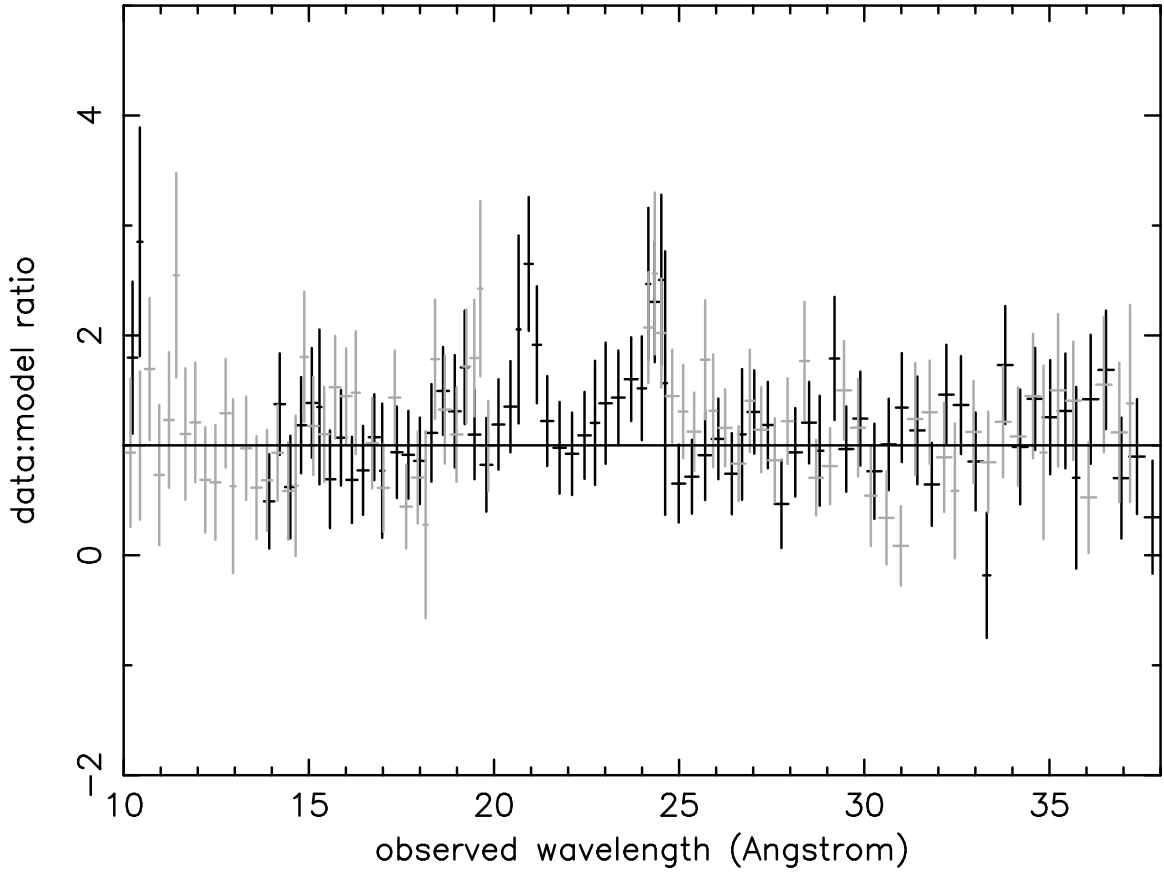}}
\caption{Ratio of RGS data to a power law and black body continuum fit revealing apparent emission lines near 21 and 24
Angstrom.\label{fig5}}
\end{figure}

\clearpage

\begin{figure}
\rotatebox{-90}{
\epsscale{0.7}
\plotone{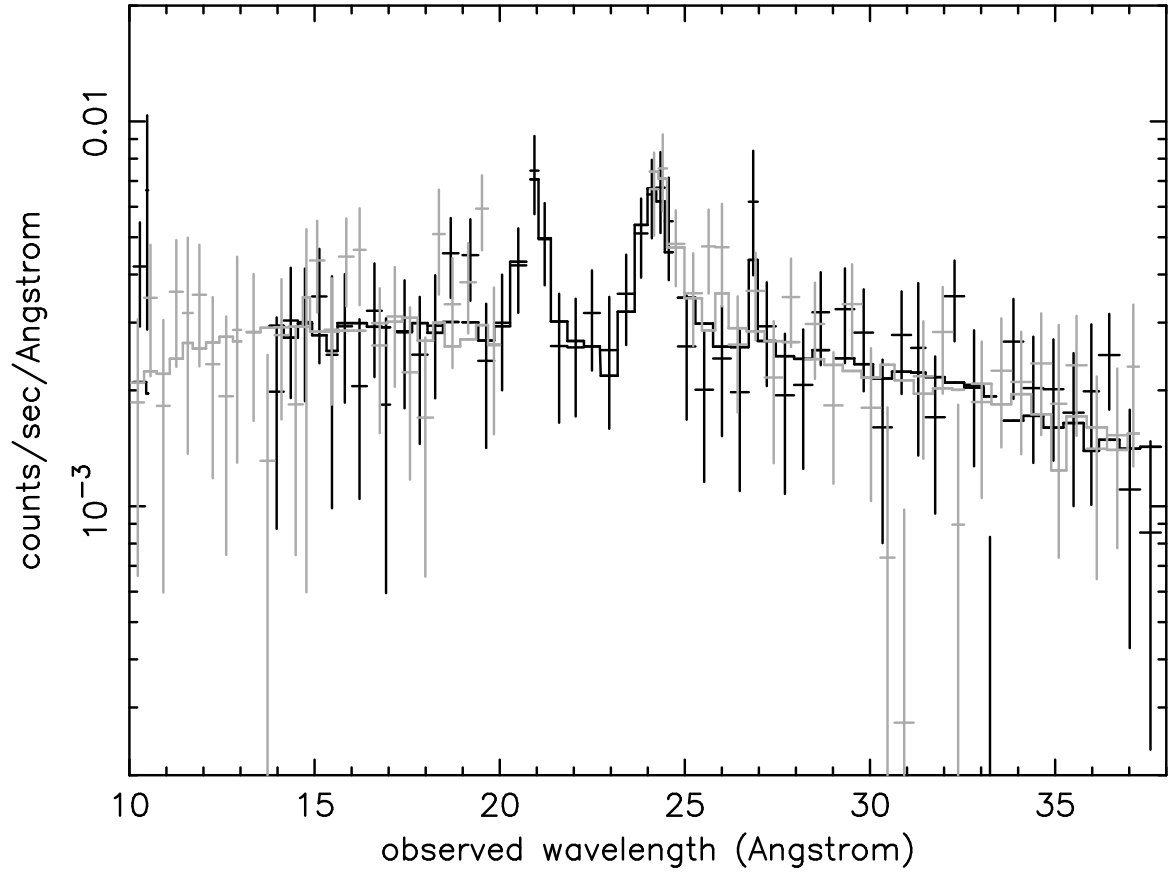}}
\caption{RGS data fitted with a power law and black body continuum plus emission lines of OVII and OVIII, as described in
Section 4.\label{fig6}}
\end{figure}

\clearpage

\begin{figure}
\rotatebox{-90}{
\epsscale{0.7}
\plotone{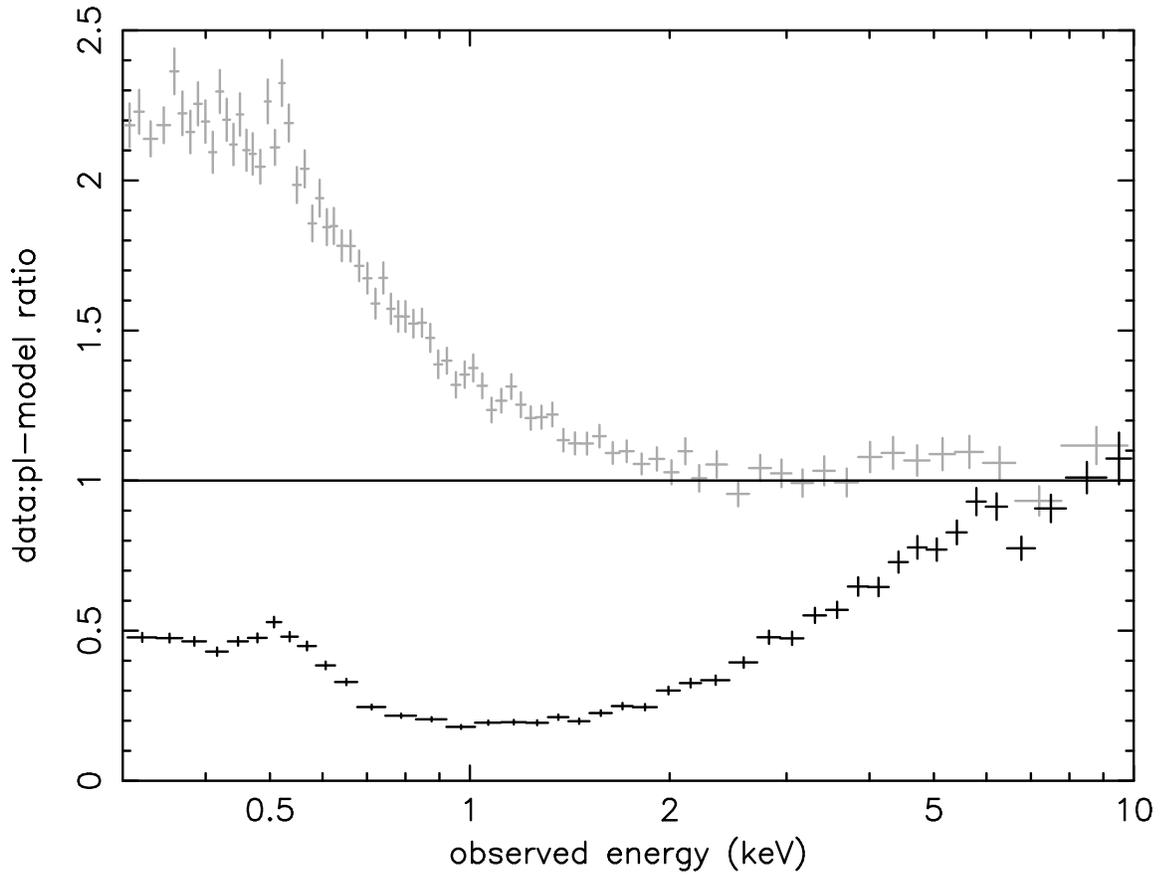}}
\caption{Ratio of EPIC pn data to 2--10 keV power law fit for the December 2000 \xmm\ observation (grey) and of the September 2002
pn data (black) to the same power law. See Section 5 for details.\label{fig7}}
\end{figure}

\clearpage

\begin{figure}
\rotatebox{-90}{
\epsscale{0.7}
\plotone{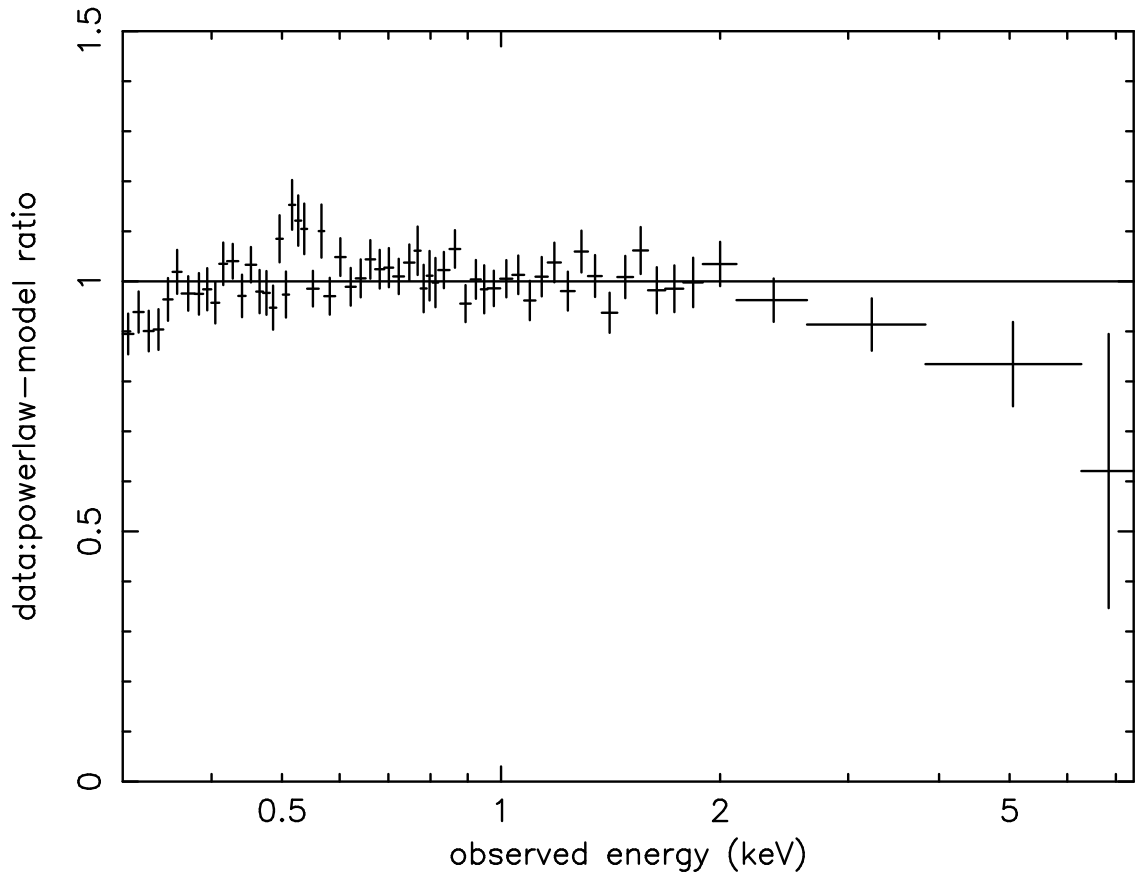}}
\caption{Simple power law fit to the difference spectrum obtained by subtraction of the present low state pn data from the high state
data obtained in December 2000. \label{fig8}}
\end{figure}

%% No more than seven \figcaption commands are allowed per page, 
%% so if you have more than seven captions, insert a \clearpage 

%% Tables should be submitted one per page, so put a \clearpage before

%% Two options are available to the author for producing tables:  the
%% deluxetable environment provided by the AASTeX package or the LaTeX
%% table environment.  Use of deluxetable is preferred.
%%

\begin{table}
\begin{center}
\caption{Laor line/PEXRAV model (parameters in AGN rest frame}
\begin{tabular}{cccccccc}
\tableline\tableline
&\\
& power-law & & \multicolumn{4}{c}{OVII/OVIII emission} \\
\cline{2-2} \cline{4-7}
instrument  & $\Gamma$ & & E (keV) & $\sigma$ (eV) & flux\tablenotemark{a} & EW (eV) & $\chi^{2}$/dof\\
\tableline
&\\
pn &  1.26$\pm$0.09 &  &   0.61$\pm$0.01 & 40$\pm$20 & 2.5$\pm$0.6 & $\sim$53 & 991/1039 \\
MOS  & 1.23$\pm$0.09 &  & 0.57$\pm$0.01 & & 2.2$\pm$0.8 & $\sim$45 \\
&\\
\tableline
&\\
&  blackbody & & \multicolumn{5}{c}{Laor line}\\
\cline{2-2} \cline{4-8}
instrument & kT (eV) & & E (keV) & $\beta$ & r$_{in}$ (r$_{g}$) & $\theta$ ($\degr$) & EW (keV)\\  
\tableline
&\\
pn+MOS & 92$\pm$2 & & 6.9$\pm$0.4 & 7.5$\pm$0.5 & 1.3$\pm$0.2 & 46$\pm$4 & 1.1$\pm$0.3\\
&\\
\tableline\tableline
\end{tabular}

%% Any table notes must follow the \end{tabular} command.

\tablenotetext{a}{10$^{-6}$ ph cm$^{-2}$ s$^{-1}$}

%\tablecomments{We can also attach a long-ish paragraph of explanatory
%material to a table. Use \tablerefs to append a list of references. (See
%the notes to the next table for an example.)}
\end{center}
\end{table}

\begin{table}
\begin{center}
\caption{Partial covering model (parameters in AGN rest frame)}
\begin{tabular}{ccccccc}
\tableline\tableline
&\\
& \multicolumn{2}{c}{power-law} & & \multicolumn{3}{c}{partial covering absorber} \\
\cline{2-3} \cline{5-7}
instrument  & $\Gamma$ & norm.\tablenotemark{a} & & norm.\tablenotemark{a} & N$_{H}$\tablenotemark{b} & $\xi$\\
\tableline
&\\
pn &  1.50$\pm$0.05 & 6.3$\pm$0.2 & & 9.6$\pm$1.4 & 4.3$\pm$0.4 & $\lesssim$0.3 \\
MOS  & 1.38$\pm$0.06 & 6.2$\pm$0.2 & & 7.2$\pm$1.4 & &  \\
&\\
\tableline
&\\
&  blackbody & & \multicolumn{4}{c}{OVII/OVIII emission}\\
\cline{2-2} \cline{4-7}
instrument & kT (eV) & & E (keV) & $\sigma$ (eV) & flux\tablenotemark{c} & EW (eV)\\  
\tableline
&\\
pn & 87$\pm$2 & & 0.61$\pm$0.01 & 48$\pm$15 & 2.9$\pm$0.7  & $\sim$65\\
MOS &  & & 0.57$\pm$0.01 & & 2.5$\pm$1.6  & $\sim$53 \\
&\\
\tableline
&\\
  &  \multicolumn{4}{c}{emission line} \\
\cline{2-5}
instrument  & E (keV) & $\sigma$ (eV) & flux\tablenotemark{c} & EW (eV) & & $\chi^{2}$/dof\\
\tableline
&\\
pn & 6.21$\pm$0.11 & 160$\pm$110 & 11$\pm$8 & $\sim$84  & & 981/1034\\
MOS  & & & 10$\pm$7  & $\sim$88 \\
&\\
\tableline\tableline
\end{tabular}

\tablenotetext{a}{10$^{-4}$ ph cm$^{-2}$ s$^{-1}$ keV$^{-1}$}
\tablenotetext{b}{10$^{22}$ cm$^{-2}$}
\tablenotetext{c}{10$^{-6}$ ph cm$^{-2}$ s$^{-1}$}

\end{center}
\end{table}

\end{document}